\documentclass[a4paper,11pt,centertags,psamsfonts]{amsart}
\usepackage{times,amssymb}

\newcommand{\Z}{\mathbb{Z}}
\newcommand{\R}{\mathbb{R}}

\newcommand{\E}{\mathbb{E}}

\newcommand{\cF}{\mathcal{F}}
\newcommand{\cJ}{\mathcal{J}}

\renewcommand{\P}{\mathbb{P}}
\newcommand{\supp}{{\ensuremath{\rm supp}}}
\newcommand{\tr}{\mathrm{tr}}

\newcommand{\sign}{\mbox{\rm sign}}

\newtheorem{theorem}{Theorem}{\bf}{\it}
\newtheorem{proposition}[theorem]{Proposition}{\bf}{\it}
{\bf}{\it}
{\it}{\rm}
\newtheorem{lemma}[theorem]{Lemma}{\bf}{\it}
{\it}{\rm}
{\bf}{\it}

\title[Lyapunov Exponent and Integrated Density
of States]{Global Bounds for the Lyapunov Exponent and the Integrated Density
of States of Random Schr\"{o}dinger Operators in One Dimension}

\author[V. Kostrykin and R. Schrader]{V. Kostrykin \and R. Schrader$^\ast$}

\address{Vadim Kostrykin\\
Fraunhofer-Institut f\"{u}r Lasertechnik\\ Steinbachstra{\ss}e 15, D-52074\\ Aachen,
Germany}
\email{kostrykin@t-online.de, kostrykin@ilt.fhg.de}

\address{Robert Schrader\\ Institut f\"{u}r
Theoretische Physik\\ Freie Universit\"{a}t Berlin, Arnimallee 14\\ D-14195 Berlin,
Germany}
\email{schrader@physik.fu-berlin.de}

\thanks{$^\ast$ R.S. supported in part by
DFG SFB 288 ``Differentialgeometrie und Quantenphysik''}

\date{May 13, 2000}



\begin{document}

\begin{abstract}
In this article we prove an upper bound for the Lyapunov exponent $\gamma(E)$
and a two-sided bound for the integrated density of states $N(E)$ at an
arbitrary energy $E>0$ of random Schr\"{o}dinger operators in one dimension. These
Schr\"{o}dinger operators are given by potentials of identical shape centered at
every lattice site but with non-overlapping supports and with randomly varying
coupling constants. Both types of bounds only involve scattering data for the
single-site potential. They show in particular that both $\gamma(E)$ and
$N(E)-\sqrt{E}/\pi$ decay at infinity at least like $1/\sqrt{E}$. As an example
we consider the random Kronig-Penney model.
\end{abstract}

\maketitle

\section{Introduction}

In this article we will consider random Schr\"{o}dinger operators $H(\omega)$ in
$L^2(\R)$ of the form
\begin{equation}\label{1.1}
H(\omega)=H_0+V_\omega,\quad H_0=-\frac{d^2}{dx^2},\quad
V_\omega=\sum_{j\in{\Z}}
\alpha_j(\omega)f(\cdot-j),
\end{equation}
where $\{\alpha_j(\omega)\}_{j\in{\Z}}$ is a sequence of i.i.d.\ (independent,
identically distributed) variables on a complete probability space
$(\Omega,\cF,\P)$ having a common distribution measure $\kappa$ (i.e.\
$\P\{\alpha_j\in\Delta\}=\kappa(\Delta)$ for any Borel set $\Delta\subset\R$).
In what follows we always suppose that $\kappa$ is supported on a compact
interval and the single-site potential $f$ is integrable with support in the
interval [-1/2,1/2]. Moreover, the random variables are assumed to form a
stationary, metrically transitive random field, i.e.\ there are measure
preserving ergodic transformations $\{T_j\}_{j\in\Z}$ such that
$\alpha_j(T_k\omega)=\alpha_{j-k}(\omega)$ for all $\omega\in\Omega$. The
spectral properties of the operator (\ref{1.1}) were studied in detail in
\cite{KiMa,Delyon,KiKoSi,Kotani:Simon:87,Sims:Stolz:2000}. The results are most
complete for the case when $f$ is the point interaction (see
\cite{Albeverio:book}).

The integrated density of states $N(E)$ and the Lyapunov exponent $\gamma(E)$
are important quantities associated with operators of the form \eqref{1.1} (see
e.g.\ \cite{Carmona:Lacroix}). In particular, according to Ishii-Pastur-Kotani
theorem \cite{Kotani} the set $\{E:\gamma(E)=0\}$ is the essential support of
the absolute continuous part of the spectral measure for $H(\omega)$.

The main idea of our approach is to approximate the operator \eqref{1.1} by
means of the sequence
\begin{equation*}
H^{(n)}(\omega)=H_0+\sum_{j=-n}^{n}
\alpha_j(\omega)f(\cdot-j)
\end{equation*}
with unchanged $H_0$, which converges to $H(\omega)$ in the strong resolvent
sense. This differs from the usual approach where one puts the whole system in
a box, which then tends to infinity (see e.g.\ \cite{Carmona:Lacroix}). In
\cite{KS1} (see also \cite{Kostrykin:Schrader:99c}) we used this approximation
to invoke scattering theory for the study the spectral properties of the
limiting operator
\eqref{1.1}. Some other applications of scattering theory to the study of
spectral properties of such type Schr\"{o}dinger operators in one dimension can be
found in \cite{KiKoSi} and \cite{Sims:Stolz:2000}.

One of the important ingredients of our approach developed in \cite{KS1} is the
Lifshitz-Krein spectral shift function. The spectral shift function naturally
replaces the eigenvalue counting function usually used to construct the density
of states for the operator \eqref{1.1}. The celebrated Birman-Krein theorem
(see e.g.\ \cite{Birman:Yafaev}) relates the spectral shift function to
scattering theory. In fact, up to a factor $-\pi^{-1}$ it may be identified
with the scattering phase for the pair ($H^{(n)}(\omega)$, $H_0$), i.e.
$\xi^{(n)}(E;\omega)=-\pi^{-1}\delta^{(n)}(E;\omega)$ when $E>0$,
\begin{displaymath}
\delta^{(n)}(E;\omega)=\frac{1}{2i}\log\det S^{(n)}(E;\omega)=
\frac{1}{2i}\log\det \left(\begin{array}{lr}
                  T_\omega^{(n)}(E) & R_\omega^{(n)}(E) \\
                  L_\omega^{(n)}(E) & T_\omega^{(n)}(E)
                  \end{array}\right).
\end{displaymath}
Here $|T^{(n)}(E)|^2$ and $|R^{(n)}(E)|^2=|L^{(n)}(E)|^2$ have the meaning of
transmission and reflection coefficients, respectively, such that
$|T^{(n)}(E)|^2+|R^{(n)}(E)|^2=1$. For $E<0$ the quantity $\xi^{(n)}(E;\omega)$
equals minus the counting function for $H^{(n)}(\omega)$.

In particular in \cite{KS1} we proved the almost sure existence of the limit
\begin{equation}\label{xi}
\xi(E)=\lim_{n\rightarrow\infty}\frac{\xi^{(n)}(E;\omega)}{2n+1},
\end{equation}
which we called the spectral shift density. Also we proved the equality
$\xi(E)=N_0(E)-N(E)$, where $N(E)$ and $N_0(E)=\pi^{-1}[\max(0,E)]^{1/2}$ are
the integrated density of states of the Hamiltonians $H(\omega)$ and $H_0$
respectively. This result also extends to higher dimension in the continuous
\cite{Kostrykin:Schrader:99d} and discrete \cite{Chahrour} cases. Also we
showed that almost surely the Lyapunov exponent $\gamma(E)$ at energy $E>0$ is
given as
\begin{equation}\label{1.3}
\gamma(E)=-\lim_{n\rightarrow\infty}\frac{\log |T^{(n)}(E;\omega)|}{2n+1},
\end{equation}
where $T^{(n)}(E,\omega)$ is the transmission amplitude for the pair of
Hamiltonians ($H^{(n)}(\omega)$, $H_{0}$) at energy $E$. We recall that
$\gamma(E)$ is defined as the upper Lyapunov exponent for the fundamental
matrix at energy $E$ of the Schr\"{o}dinger operator $H(\omega)$. The connection
between the Lyapunov exponent and the transmission coefficient
$|T_\omega^{(n)}(E)|$ was recognized long ago
\cite{Lifshitz:Gredeskul:Pastur:82,Lifshitz:Gredeskul:Pastur:88}. A complete
proof has appeared in \cite{KS1}.

We note that the theory of the spectral shift function was also recently used
to show that the integrated density of states is independent of the choice of
boundary conditions \cite{Nakamura:2000} on the sides of a large box, in which
the system is put.

The conditions on the random variables $\alpha_j$ and the single-site potential
$f$ stated above are slightly weaker than those in \cite{KS1}. However the
results of \cite{KS1} which will be used below remain valid also in this more
general case.

The aim of the present paper is to prove \emph{global} bounds for the Lyapunov
exponent and the integrated density of states, i.e.\ bounds which hold for all
$E>0$ and describe the correct asymptotic behavior in the limit
$E\rightarrow\infty$. These results are formulated as Theorems \ref{theorem1}
and \ref{thm:xi} below. To the best of our knowledge the first article to look
for the asymptotic behavior of $\gamma(E)$ and $N(E)$ in the limit
$E\rightarrow\infty$ is \cite{APW}. The best known estimate for the integrated
density of states is due to Kirsch and Martinelli \cite[Corollary 3.1]{KiMa2}.
This bound however does not reproduce the correct asymptotic behavior of $N(E)$
in the large energy limit. Another estimate, which is due to Pastur and Figotin
(see \cite[Sec.\ V.11.B]{Pastur:Figotin}), is valid for an $\R$-metrically
transitive random field. Since our potential $V_\omega(x)$ is a $\Z$-metrically
transitive field this estimate does not apply directly to the present
situation. Our two-sided estimate leads to the bound \eqref{IDS} below which is
very close to that of Pastur and Figotin.

In what follows $C$ will denote a finite positive generic constant varying with
the context, but which depends only on $f$ and $\kappa$.

We are indebted to Leonid Pastur for reading the preliminary version of this
article.

\section{The Lyapunov exponent}

We recall that the scattering matrix $S(E)$ for a pair of Hamiltonians ($H$,
$H_0$) on $L^2(\R)$ at fixed energy $E\geq 0$ is a $2\times 2$ unitary matrix
\begin{equation}\label{Smatrix.def}
S(E)=\left(\begin{array}{cc}
           T(E) & R(E) \\
           L(E) & T(E)
           \end{array} \right),
\end{equation}
where $L(E)$ and $R(E)$ denote the left and right reflection amplitudes
respectively. The transmission amplitude $T(E)$ can vanish only for $E=0$ (see
\cite{Faddeev,Deift:Trubowitz}). To any S-matrix \eqref{Smatrix.def} we
associate the unimodular matrix
\begin{equation*}
\Lambda(E)=\begin{pmatrix}
\frac{1}{T(E)} & -\frac{R(E)}{T(E)}\\[2mm]
\frac{L(E)}{T(E)} & \frac{1}{\ \overline{T(E)}\ }
\end{pmatrix}.
\end{equation*}
Let $T_\alpha(E)$, $R_\alpha(E)$, $L_\alpha(E)$ be the elements of the S-matrix
at energy $E$ for the pair of operators ($H_0+\alpha f$, $H_0$) and
$\Lambda_{\alpha}(E)$ the corresponding $\Lambda$-matrix. Also let
$\widetilde{\Lambda}_{\alpha}(E)=U_{E}^{1/2}\Lambda_{\alpha}(E)U^{1/2}_{E}$
with
\begin{equation*}
U_E=\begin{pmatrix} e^{i\sqrt{E}} & 0 \\
0 & e^{-i\sqrt{E}}
\end{pmatrix}.
\end{equation*}
Explicitly we have
\begin{equation*}
\widetilde{\Lambda}_{\alpha}(E)=\begin{pmatrix}
\frac{e^{i\sqrt{E}}}{T_\alpha(E)} &
-\frac{R_\alpha(E)}{T_\alpha(E)}\\[2mm]
\frac{L_\alpha(E)}{T_\alpha(E)} &
\frac{e^{-i\sqrt{E}}}{\ \overline{T(E)}\ }
\end{pmatrix}.
\end{equation*}

Consider the matrix
\begin{displaymath}
A(E)=\E\left\{\widetilde{\Lambda}_{\alpha(\omega)}(E)^{\dagger}
\widetilde{\Lambda}_{\alpha(\omega)}(E) \right\}=\int \widetilde{\Lambda}_{\alpha}(E)^{\dagger}
\widetilde{\Lambda}_{\alpha}(E)d\kappa(\alpha)\geq 0,
\end{displaymath}
where for brevity we write $\alpha(\omega)$ instead of $\alpha_j(\omega)$ with
some $j\in\Z$. Let $\beta_{+}(E)$ be the largest eigenvalue of $A(E)$ and
$\beta_{-}(E)$ the smallest. It will turn out below that $\beta_{+}(E)\ge 1$.
Set $\widetilde{\gamma}(E)=(\log
\beta_{+}(E))/2\geq 0$.

The first main result of the present article is

\begin{theorem}\label{theorem1}
Given the Hamiltonian \eqref{1.1} and the distribution $\kappa$ for the
coupling constant $\alpha$, for all $E>0$ the resulting Lyapunov exponent
satisfies the upper bound
\begin{equation}\label{1.4}
\gamma(E)\le\widetilde{\gamma}(E).
\end{equation}
In particular $\gamma(E)$ decays at least like $1/\sqrt{E}$ at infinity.
\end{theorem}

\begin{proof}
Let $\Lambda^{(n)}(E;\omega)$ denote the $\Lambda$-matrix for the pair
($H^{(n)}(\omega)$, $H_{0}$), which by the factorization property can be
represented in the form
\begin{equation}\label{Lambda.U}
\Lambda^{(n)}(E;\omega)=U_E^{-n-1/2}\prod_{j=-n}^n
\widetilde{\Lambda}_{\alpha_j(\omega)}(E)\cdot U_E^{-n-1/2}.
\end{equation}
In fact, this factorization property is a consequence of the multiplicativity
property of the fundamental matrix (see \cite{KS1} for a proof and for
references to earlier work). A short calculation gives
\begin{equation}\label{1.5}
|T^{(n)}(E;\omega)|^{-2}=\frac{1}{4}
\tr\left(\Lambda^{(n)}(E;\omega)^{\dagger}\Lambda^{(n)}(E;\omega)\right)
+\frac{1}{2}.
\end{equation}
With $\E$ denoting the expectation with respect to the measure $\P$, by
Jensen's inequality and \eqref{1.5} we therefore have the estimate
\begin{eqnarray}\label{1.6}
\lefteqn{e^{-2\E\{\log |T^{(n)}(E;\omega)|\}}\le
\E\left\{|T^{(n)}(E;\omega)|^{-2}\right\}}\nonumber\\ &=& \frac{1}{4}
\E\left\{\tr\left(\Lambda^{(n)}(E;\omega)^{\dagger}
\Lambda^{(n)}(E;\omega)\right)\right\}+\frac{1}{2}.
\end{eqnarray}
From the factorization property \eqref{Lambda.U} it follows that
\begin{equation}\label{1.7}
\tr\left(\Lambda^{(n)}(E;\omega)^{\dagger}\Lambda^{(n)}(E;\omega)\right)
=\tr\left(\prod_{j=n}^{-n}
\widetilde{\Lambda}_{\alpha_j(\omega)}(E)^{\dagger}\prod_{j=-n}^n
\widetilde{\Lambda}_{\alpha_j(\omega)}(E)\right).
\end{equation}
We will now make use of the fact that the $\alpha_{k}(\omega)$ are i.i.d.\
random variables. For this purpose define the $2\times 2$ matrices
$A_{j}(E)\geq 0$ recursively by $A_{0}=I$ and
\begin{equation}\label{1.8}
A_{j}(E)=\int \widetilde{\Lambda}_{\alpha}(E)^{\dagger}A_{j-1}(E)
\widetilde{\Lambda}_{\alpha}(E)d\kappa(\alpha),
\end{equation}
such that in particular $A(E)=A_{1}(E)$.
Now it is easy to see that
\begin{eqnarray}\label{1.10}
\lefteqn{\E\left\{\tr\left(\Lambda^{(n)}(E;\omega)^{\dagger}
\Lambda^{(n)}(E;\omega)\right)\right\}}\nonumber\\ &=&
\tr\left(\E\left(\Lambda^{(n)}(E;\omega)^{\dagger}
\Lambda^{(n)}(E;\omega)\right)\right)
=A_{2n+1}(E).
\end{eqnarray}
We now use the fact that the operator inequality $0\le A\le A^{\prime}$ implies
$0\le \tr\,A\le\tr\,A^{\prime}$ and $B^{\dagger}AB\le B^{\dagger}A^{\prime}B$
for all $B$. In particular we have $A(E)\le \beta_{+}(E)\,I$ from which we
obtain the recursive estimates $A_{j}(E)\le
\beta_{+}(E)A_{j-1}(E)\le\cdots\le\beta_{+}(E)^{j}\, I$ and hence
\begin{equation}\label{1.11}
\E\left\{\tr\left(\Lambda^{(n)}(E;\omega)^{\dagger}
\Lambda^{(n)}(E;\omega)\right)\right\}\le 2\beta_{+}(E)^{2n+1}.
\end{equation}
We remark that with the same arguments one proves the lower bound
\begin{equation*}
2\beta_{-}(E)^{2n+1}\le\E\left(\tr\left(\Lambda^{(n)}(E;\omega)^{\dagger}
\Lambda^{(n)}(E;\omega)\right)\right).
\end{equation*}
The relation \eqref{1.3}, the estimate \eqref{1.11} combined with \eqref{1.6}
and Fatou's lemma imply now
\begin{eqnarray*}
\gamma(E) &\leq & \frac{1}{2}\lim_{n\rightarrow\infty}
\frac{\log \E\left\{|T^{(n)}(E;\omega)|^{-2}\right\}}{2n+1}\\
&\leq & \frac{1}{2}\lim_{n\rightarrow\infty}\frac{\log\left(\beta_+(E)^{2n+1}/2
+1/2\right)}{2n+1}=\frac{1}{2}\log \beta_+(E),
\end{eqnarray*}
which proves the claim \eqref{1.4}.

To establish the last
claim of the theorem we recall the following well known estimates (see e.g.\
\cite{Faddeev,Deift:Trubowitz})
\begin{equation}\label{est}
|T_{\alpha}(E)-1|+|R_{\alpha}(E)|\le C\frac{1}{\sqrt{E}}
\end{equation}
valid for all large $E>0$ uniformly for all $\alpha$ in the (compact) support
of $\kappa$ for fixed $f$. Using the estimate \eqref{est} in \eqref{1.12} gives
the estimate $\beta_{+}(E)\le 1+C/\sqrt{E}$ for all large $E$. Since
$\widetilde{\gamma}(E)=(\log\beta_{+}(E))/2$, this concludes the proof of the
theorem.
\end{proof}

Since $\gamma(E)\ge 0$, we obviously have the inequality $\beta_{+}(E)\ge 1$
for almost all $E$. We will give now a direct independent proof of this fact
and simultaneously obtain an expression for $\beta_{+}(E)$. The matrix $A(E)$
may be written in the form
\begin{displaymath}
A(E)=\begin{pmatrix}
a(E)& b(E) \\
\overline{b(E)}& a(E)
\end{pmatrix}
\end{displaymath}
with
\begin{eqnarray}\label{1.12}
 a(E)&=&\int\left(\frac{2}{|T_{\alpha}(E)|^{2}}-1\right)d\kappa(\alpha)\\
 b(E)&=&-e^{i\sqrt{E}}\int\frac{R_{\alpha}(E)}{T_{\alpha}(E)^{2}}
d\kappa(\alpha).
\end{eqnarray}
This gives the two eigenvalues of $A(E)$ in the form
\begin{equation}
\label{1.13}
\beta_{\pm}(E)=a(E)\pm |b(E)|
\end{equation}
Obviously $a(E)\ge 1$ and hence $\beta_{+}(E)\ge 1$. In fact, $a(E)=1$ is
possible if and only if $R_{\alpha}(E)=0$ for almost all $\alpha$ in the
support of $\kappa$. Then also $b(E)=0$ and $\beta_{+}(E)=1$. Actually (if
$\supp\ \kappa$ has at least one non-isolated point) we do not believe there
are nontrivial $f$ and $E$ for which this holds but in any case for such $E$'s
the Lyapunov exponent vanishes as is easily verified (see also \cite{KS1}), so
this is a trivial confirmation of estimate
\eqref{1.4} in this case. In the remaining case we trivially have $\beta_{+}(E)>1$.

As an example we consider the random Kronig - Penney model which is formally
obtained from $H(\omega)$ by replacing $f$ with the Dirac $\delta$-function at
the origin. Then we have (correcting for a misprint on page 232 of \cite{KS1})
\begin{eqnarray}
\label{1.14}
T_{\alpha}(E)&=&\left(1+i\frac{\alpha}{2\sqrt{E}}\right)^{-1}\\
R_{\alpha}(E)&=&
-i\frac{\alpha}{2\sqrt{E}}\left(1+i\frac{\alpha}{2\sqrt{E}}\right)^{-1}
\end{eqnarray}
and our method still applies.
This gives
\begin{eqnarray}
\label{1.15}
a(E)&=&1+\frac{\langle\alpha^{2}\rangle}{4E}\\
b(E)&=&i\frac{\langle\alpha\rangle}{2\sqrt{E}}-\frac{\langle\alpha^{2}\rangle}{4E}.
\end{eqnarray}
Here for brevity by $\langle\;\rangle$ we denote the mean with respect to the
probability measure $\P$ such that
\begin{equation*}
\langle\alpha\rangle = \E\{\alpha(\omega)\} = \int\alpha d\kappa(\alpha),\qquad
\langle\alpha^2\rangle = \E\{\alpha(\omega)^2\} = \int\alpha^2 d\kappa(\alpha).
\end{equation*}
In particular
\eqref{1.15} gives
\begin{equation}
\label{1.16}
\beta_{+}(E)=1+\frac{\langle\alpha^{2}\rangle}{4E}+\frac{1}{2\sqrt{E}}
\left(\frac{\langle\alpha^{2}\rangle}{4E}+\langle\alpha\rangle^{2}\right)^{1/2}.
\end{equation}
So also in this case $\gamma(E)$ decays at least like $1/\sqrt{E}$ as
$E\rightarrow\infty$ and at least like $1/E$ if the mean $\langle\alpha\rangle$
of $\alpha$ vanishes, i.e.\ if on average the coupling constant is zero.

\section{The integrated density of states}

We denote by $\xi_\alpha(E)$ the spectral shift function for the pair
$(H_0+\alpha f, H_0)$. The second main result of this article is given by

\begin{theorem}\label{thm:xi}
For all $E>0$ the spectral shift density $\xi(E)$ for the operator \eqref{1.1}
satisfies the following two-side bound
\begin{equation}\label{esti}
\E\{\xi_{\alpha(\omega)}(E)\}-r(E)\leq \xi(E)\leq
\E\{\xi_{\alpha(\omega)}(E)\}+r(E),
\end{equation}
where
\begin{equation*}
r(E)=\min\left\{\frac{1}{2},
\frac{1}{\pi}\E\left\{\frac{|R_{\alpha(\omega)}(E)}{1-|R_{\alpha(\omega)}(E)|}\right\}\right\}.
\end{equation*}
In particular $\E\{\xi_{\alpha(\omega)}(E)\}$ and $r(E)$ decays at least like
$1/\sqrt{E}$ at infinity.
\end{theorem}

\textit{Remarks}.
1. One can easily prove the following estimate
\begin{equation*}
\E\{\xi_{\alpha(\omega)}(E)\}-1\leq \xi(E)\leq
\E\{\xi_{\alpha(\omega)}(E)\}+1,
\end{equation*}
which is valid for all $E\in\R$.

2. By the monotonicity of the spectral shift function with respect to
perturbation $\xi(E)\geq 0$ if $\supp\ \kappa\subset\R_+$ and $\xi(E)\leq 0$ if
$\supp\ \kappa\subset\R_-$ for almost all $E>0$.

3. For large $E>0$ by \eqref{est}
\begin{equation*}
r(E)=\min\left\{\frac{1}{2},
\frac{1}{\pi}\E\left\{\frac{|R_{\alpha(\omega)}(E)}{1-|R_{\alpha(\omega)}(E)|}\right\}\right\}
=\frac{1}{\pi}\E\left\{\frac{|R_{\alpha(\omega)}(E)}{1-|R_{\alpha(\omega)}(E)|}\right\}
\leq \frac{C}{\sqrt{E}}.
\end{equation*}

4. In \cite{KS1} we proved the relation $\xi(E)=N_0(E)-N(E)=\sqrt{E}/\pi-N(E)$,
where $N_0(E)$ is the integrated density of states for the free operator $H_0$.
Theorem \ref{thm:xi} then gives the following two-sided bound for the
integrated density of states
\begin{equation}\label{IDS}
\frac{\sqrt{E}}{\pi}-\E\{\xi_{\alpha(\omega)}(E)\}-r(E)\leq N(E)
\leq \frac{\sqrt{E}}{\pi}-\E\{\xi_{\alpha(\omega)}(E)\}+r(E),\quad E>0.
\end{equation}
There are some other upper bounds on the integrated density of states. A
well-known result is a one-sided bound due to Kirsch and Martinelli
\cite[Corollary 3.1]{KiMa2},
\begin{equation*}
N(E)\leq \frac{C}{\sqrt{\eta}}\
\E\left\{\int_{-1/2}^{1/2}(E+\eta-V_\omega(x))_+ dx
\right\}
\end{equation*}
for any $\eta>0$ and all $E\in\R$. This bound however does not reproduce the
correct asymptotic behavior of $N(E)$ in the large energy limit.

5. The bounds \eqref{1.4} and \eqref{esti} are of interest in the context of
the Thouless formula (see e.g.\ \cite{Pastur:Figotin})
\begin{equation}\label{Thouless}
\gamma(E)-\gamma_0(E)=-\int_\R \log|E-E^\prime|\ d\xi(E^\prime),\qquad E\in\R,
\end{equation}
where $\gamma_0(E)=[\max(0,-E)]^{1/2}$ is the Lyapunov exponent for $H_0$. The
Thouless formula in the form \eqref{Thouless} can be viewed as a subtracted
dispersion relation (see e.g.\ \cite{KS1}).

\begin{proof}
In \cite{KS1} we proved (see Theorem 3.3 there and its proof) that for any two
potentials $V_1$ and $V_2$ with (compact) disjoint supports one has
\begin{equation*}
\xi(E;H_0+V_1+V_2,H_0) = \xi(E;H_0+V_1,H_0)+\xi(E;H_0+V_2,H_0)+\xi_{12}(E)
\end{equation*}
with
\begin{equation*}
\xi_{12}(E)=-\frac{1}{2\pi i}\log\frac{1-R_1(E)\ L_2(E)}{1-\overline{R_1(E)}\
\overline{L_2(E)}},
\end{equation*}
where $R_k(E)$ and $L_k(E)$ are the right and left reflection coefficients for
the Schr\"{o}dinger equation with the potential $V_k$, $k=1,2$. Actually Theorem
3.3 in \cite{KS1} states that $|\xi_{12}(E)|\leq 1/2$ for all $E\geq 0$. Now we
improve on this estimate. As in \cite{KS1} we set
\begin{displaymath}
L_k(E)=a_k(E)e^{i\delta_k^{(L)}},\quad R_k(E)=a_k(E)e^{i\delta_k^{(R)}},\quad
k=1,2
\end{displaymath}
with $0\leq a_k(E)\leq 1$. Moreover $a_k(E)=1$ only when $T_k(E)=0$, which we
recall can happen only if $E=0$. Therefore
\begin{eqnarray*}
&&\log\frac{1-R_1(E)\ L_2(E)}{1-\overline{R_1(E)}\
\overline{L_2(E)}}\\
&&=\log\frac{1-a_1(E)a_2(E)e^{i(\delta_1^{(R)}+\delta_2^{(L)})}}
{1-a_1(E)a_2(E)e^{-i(\delta_1^{(R)}+\delta_2^{(L)})}}\\
&&=-2i\arctan\frac{a_1(E)a_2(E)\sin(\delta_1^{(R)}
+\delta_2^{(L)})}{1-a_1(E)a_2(E)\cos(\delta_1^{(R)}+\delta_2^{(L)})}.
\end{eqnarray*}
By means of the inequality $|\arctan x| \leq |x|$ we immediately obtain
\begin{equation}\label{ineq}
|\xi_{12}(E)|\leq \min\left\{\frac{1}{2}, \frac{1}{\pi}\frac{a_1(E)
a_2(E)}{1-a_1(E) a_2(E)}\right\}.
\end{equation}
Since $0\leq a_k(E)<1$ we can replace $a_1(E) a_2(E)(1-a_1(E) a_2(E))^{-1}$
either by $a_1(E)(1-a_1(E))^{-1}$ or by $a_2(E)(1-a_2(E))^{-1}$.

Now let us consider the operator $H^{(n)}(\omega)$ for finite $n$. Applying the
inequality \eqref{ineq} we obtain
\begin{eqnarray*}
\lefteqn{\left|\xi^{(n)}(E;\omega)-\xi_{\alpha_n(\omega)}(E)-\xi_{\alpha_{-n}(\omega)}(E)
-\xi^{(n-1)}(E;\omega) \right|}\\
&\leq& \min\left\{\frac{1}{2}, \frac{1}{\pi}\frac{|R_{\alpha_n(\omega)}(E)|}
{1-|R_{\alpha_n(\omega)}(E)|} \right\}+
\min\left\{\frac{1}{2}, \frac{1}{\pi}\frac{|R_{\alpha_{-n}(\omega)}(E)|}
{1-|R_{\alpha_{-n}(\omega)}(E)|} \right\}
\end{eqnarray*}
Repeating this procedure recursively we obtain
\begin{displaymath}
\left|\xi^{(n)}(E;\omega)-\sum_{j=-n}^{n} \xi_{\alpha_j(\omega)}(E)\right|\leq
\sum_{j=-n}^n \min\left\{\frac{1}{2},
\frac{1}{\pi}\frac{|R_{\alpha_j(\omega)}(E)|}
{1-|R_{\alpha_j(\omega)}(E)|}\right\}.
\end{displaymath}
From the existence of the spectral shift density \eqref{xi} by the Birkhoff
ergodic theorem it follows that
\begin{equation*}
\left|\xi(E)-\E\left\{\xi_{\alpha(\omega)}(E)\right\}\right|\leq \E\left\{\min\left\{\frac{1}{2},
\frac{1}{\pi}\frac{|R_{\alpha(\omega)}(E)|}
{1-|R_{\alpha(\omega)}(E)|}\right\} \right\}.
\end{equation*}
From the obvious inequality
\begin{equation*}
\E\left\{\min\left\{\frac{1}{2},
\frac{1}{\pi}\frac{|R_{\alpha(\omega)}(E)|}
{1-|R_{\alpha(\omega)}(E)|}\right\} \right\}\leq
\min\left\{\frac{1}{2},\frac{1}{\pi}\E\left\{\frac{|R_{\alpha(\omega)}(E)|}
{1-|R_{\alpha(\omega)}(E)|}\right\} \right\}
\end{equation*}
the bound \eqref{esti} follows.

For large $E$ we have the following asymptotics \cite{Deift:Trubowitz}
uniformly in $\alpha$ on compact sets:
\begin{eqnarray*}
R_{\alpha}(E) &=& \frac{\alpha}{2i\sqrt{E}}\int_\R
e^{2i\sqrt{E}t}f(t)dt+O(E^{-1}),\\ L_{\alpha}(E) &=&
\frac{\alpha}{2i\sqrt{E}}\int_\R e^{-2i\sqrt{E}t}f(t)dt+O(E^{-1})
\end{eqnarray*}
such that $R_{\alpha}(E)=O(1/\sqrt{E})$ and $L_{\alpha}(E)=O(1/\sqrt{E})$. If
the single-site potential $f$ has $p$ derivatives in $L^1(\R)$ then
$L_{\alpha}(E)=O(E^{-(p+1)/2})$ and $R_{\alpha}(E)=O(E^{-(p+1)/2})$ as
$E\rightarrow\infty$ \cite{Deift:Trubowitz}. The estimate
$\E\{\xi_{\alpha(\omega)}(E)\}=O(1/\sqrt{E})$ is Proposition \ref{prop3} below.
\end{proof}

As an example we consider again the random Kronig-Penney model. The single-site
spectral shift function is given in this case by
\begin{equation*}
\xi_\alpha(E)=\frac{1}{\pi}\arctan\left(\frac{\alpha}{2\sqrt{E}}\right),\quad E>0.
\end{equation*}
Therefore
\begin{equation*}
\E\left\{\xi_{\alpha(\omega)}(E)\right\}=
\frac{1}{\pi}\int_\R \arctan\left(\frac{\alpha}{2\sqrt{E}}\right)d\kappa(\alpha)
\end{equation*}
and thus
\begin{equation*}
\left|\E\left\{\xi_{\alpha(\omega)}(E)\right\}\right|\leq
\frac{\langle|\alpha|\rangle}{2\pi\sqrt{E}}.
\end{equation*}
Using the explicit expression for the reflection amplitude one can easily show
that
\begin{equation*}
\frac{\langle|\alpha|\rangle}{2\sqrt{E}}+\frac{\langle\alpha^2\rangle}{4E}\leq
\E\left\{\frac{|R_{\alpha(\omega)}(E)|}{1-|R_{\alpha(\omega)}(E)|}\right\}
\leq \frac{\langle|\alpha|\rangle}{2\sqrt{E}}+\frac{\langle\alpha^2\rangle}{2E}.
\end{equation*}

We complete this section with an estimate on $\E\{\xi_{\alpha(\omega)}(E)\}$
in the general case. We will prove

\begin{proposition}\label{prop3}
There is a constant $c>0$ independent of $E$, $f$, and $\kappa$ such that for all $E>0$
\begin{equation*}
\left|\E\left\{\xi_{\alpha(\omega)}(E)\right\}\right|\leq \frac{C}{2\sqrt{E}}
\E\left\{|\alpha(\omega)|^{1/2}\right\}^2 \int_{-1/2}^{1/2}|f(x)|dx.
\end{equation*}
\end{proposition}

Let $l^{1/2}(L^1)$ denote the Birman-Solomyak class of measurable functions $V$
for which
\begin{equation*}
\|V\|_{l^{1/2}(L^1)}=\left[\sum_{j=-\infty}^\infty
\left(\int_{j-1/2}^{j+1/2}|V(x)|dx \right)^{1/2} \right]^{2}<\infty.
\end{equation*}
The claim of the proposition immediately follows from the following

\begin{lemma}
Let $V\in l^{1/2}(L^1)$. There is a constant $c_1$ independent of $V$ and $E$
such that
\begin{equation*}
|\xi(E;H_0+V,H_0)|\leq \frac{c_1}{2\sqrt{E}}\|V\|_{l^{1/2}(L^1)}
\end{equation*}
for all $E>0$.
\end{lemma}

\begin{proof}
As proved in \cite{Sobolev} there is a constant $c_2>0$ independent of $E$ and
$V$ such that
\begin{equation*}
|\xi(E; H_0+V,H_0)|\leq C_1 \|V^{1/2} R_0(E+i0)|V|^{1/2}\|_{\cJ_1},
\end{equation*}
where $V^{1/2}=\sign V\ |V|^{1/2}$, $R_0(z)=(H_0-z)^{-1}$, and
$\|\cdot\|_{\cJ_1}$ denotes the trace class norm (see e.g.\ \cite{Simon:79a}).
From the proof of Proposition 5.6 in \cite{Simon:79a} it follows that
\begin{equation*}
\|V^{1/2} R_0(E+i0)|V|^{1/2}\|_{\cJ_1}\leq \frac{c_3}{\sqrt{E}}\|V\|_{l^{1/2}(L^1)}
\end{equation*}
for all $E>0$.
\end{proof}


\end{document}